\documentclass[twocolumn,aps,showpacs,superscriptaddress,tightenlines,a4paper]{revtex4}

\usepackage{amssymb}
\usepackage{amsmath}
\usepackage{bbold}
\usepackage{mathrsfs}
\usepackage{paralist}
\usepackage{graphicx}
\usepackage{xcolor}

\begin{document}
%
%
	\title{On the consistency of certain constitutive relations with quantum electromagnetism}
%
%
    \author{S. A. R. Horsley}
    \affiliation{School of Physics and Astronomy, University of St Andrews,
North Haugh, St Andrews, KY16 9SS, UK}
    \email{sarh@st-andrews.ac.uk}
%
%
    \begin{abstract}
    	Recent work by T. G. Philbin~\cite{philbin2010} has provided a Lagrangian theory that establishes a general method for the canonical quantization of the electromagnetic field in any dispersive, lossy, linear dielectric.  Working from this theory, we extend the Lagrangian description to reciprocal and non--reciprocal magnetoelectric (bi--anisotropic) media, showing that some versions of the constitutive relations are inconsistent with a real Lagrangian, and hence with quantization.  This amounts to a restriction on the magnitude of the magnetoelectric coupling.  Moreover from the point of view of quantization, moving media are shown to be fundamentally different from stationary magnetoelectrics, despite the formal similarity in the constitutive relations.
    \end{abstract}
%
%
    \maketitle
%
%
%
%
	\section{Introduction}
	\par
	In classical electromagnetism life is made much simpler by the introduction of \(\boldsymbol{\epsilon}\) \& \(\boldsymbol{\mu}\).  The microscopic current and charge densities can be forgotten, and in place of these we can deal with the macroscopic Maxwell equations~\cite{volume8}.  Indeed, for the purposes of further speeding up calculations, classical electromagnetism allows us to make artificial simplifications, such as the existence of media without dispersion, or loss.
	\par
	Quantum mechanics does not take to this description so easily.  Although the quantization of the free electromagnetic field can be found in standard textbooks~\cite{volume4}, there has been some historical difficulty in quantizing the electromagnetic field within a dielectric medium.  The field can be quantised in a fictional medium without dispersion or loss~\cite{jauch1948}, but as soon as dispersion is introduced, the procedure becomes awkward~\cite{drummond1990}.  Moreover, it was not immediately obvious how an effective description of loss might be implemented in quantum mechanics.  The classical field amplitudes ought to decay, whereas their operator counterparts must satisfy the canonical commutation relations uniformly throughout space, at all times.
	\par
	In fact these apparently distinct difficulties have their origins within a single physical effect.  Dispersion and loss are the two sides of one phenomenon; the finite response time of a material to events that happened in the past.  It is therefore through re--introducing some degrees of freedom associated with the medium that quantization may be carried out.
	\par
	Canonical quantization was achieved for a model Lagrangian by Huttner and Barnett~\cite{huttner1992}, who introduced a bath of harmonic oscillators to account for the dynamics of a uniform, dispersive and lossy dielectric.  Subsequently, this model was extended to non--uniform dielectrics~\cite{suttorp2004,suttorp2004b}, and more recently, it was recognised that some model aspects of the theory could be removed, and that the theory could describe general features of non--isotropic magnetodielectrics~\cite{amooshahi2008,kheirandish2008,amooshahi2009b}, and even moving media~\cite{amooshahi2009a}.
	\par
	Most recently, a Lagrangian density was found in~\cite{philbin2010} that describes the electromagnetic field within any linear magnetodielectric that satisfies the Kramers--Kronig relations.  From this Lagrangian a Hamiltonian was derived that allowed a canonical quantization of the electromagnetic field, and from this, formerly phenomenological results, such as the theory of Casimir forces (Lifshitz theory) have been given a canonical basis~\cite{philbin2011}.
	\par
	The approach here is to take seriously the Lagrangian in~\cite{philbin2010}.  This Lagrangian represents an arbitrary linear material that automatically satisfies the Kramers--Kronig relations, as well as some general properties usually arising from thermodynamics (see section \ref{summary-section}). On top of this, the theory can be quantized, which suggests that this may be a more fundamental, and correct way to describe macroscopic electromagnetism: the material degrees of freedom have returned, but only in the most minimal way.
	\par
	We examine extensions to the Lagrangian that describe the effects of magnetoelectric coupling, as well as time irreversibility (e.g. a medium in an external magnetic field).  Here we understand magnetoelectric materials to include chiral media, moving media, and any other media where the constitutive relations are of the form, \(\tilde{\boldsymbol{D}}=\boldsymbol{\epsilon}\boldsymbol{\cdot}\tilde{\boldsymbol{E}}+\boldsymbol{\chi}_{\text{\tiny{E\,B}}}\boldsymbol{\cdot}\tilde{\boldsymbol{B}}\), and \(\tilde{\boldsymbol{H}}=\boldsymbol{\mu}^{-1}\boldsymbol{\cdot}\tilde{\boldsymbol{B}}-\boldsymbol{\chi}_{\text{\tiny{B\,E}}}\boldsymbol{\cdot}\tilde{\boldsymbol{E}}\).  Throughout what follows, a tilde over a vector denotes it being in the frequency domain.
	\par
	The aim is so that; (a) we may understand the origin of these various effects in terms of interaction terms within a Lagrangian; and (b) we may ask whether the requirement of the existence of a corresponding Hamiltonian (from which we may quantize the field) places restrictions on the parameters within the constitutive relations.  We find the answer to (b) is positive, a result which may clarify the apparent confusion over the restrictions placed on magnetoelectric media~\footnote{Several separate inequivalent inequalities for the magnetoelectric susceptibility appear in the literature, containing the permittivity and permeability~\cite{odell1963}, the susceptibilities~\cite{brown1968}, and the \emph{imaginary} parts of the permittivity and permeability~\cite{sihvola2001}.  It does not seem to be agreed which of these is correct, and here we propose that the restriction involving the imaginary parts is the most fundamental.  Moreover, most often no distinction is made between the reciprocal magneoelectrics (e.g. chiral media), and the non--reciprocal ones (e.g. Tellegen media), for example see~\cite{shuvaev2010}.  We also note that it is pointed out in~\cite{zhang2007} that, for a fixed frequency, there is no obvious restriction on the real parts of the chiral parameters.} (see~\cite{odell1963,brown1968,sihvola2001,zhang2007}).  This confusion between the various proposed restrictions is not negligible, and is important for metamaterial design~\cite{tretyakov2003,pendry2004}, as well as a possible route to a repulsive Casimir effect~\cite{zhao2009}.
%
%
	\section{A summary of the Lagrangian theory of macroscopic electromagnetism\label{summary-section}}
	\par
	We begin with a brief review of the basic features of the Lagrangian theory of macroscopic electromagnetism presented in~\cite{philbin2010}.  It is worth emphasizing that here we are concerned with the existence of a Hamiltonian that can be used to describe the interaction of electromagnetism with more general materials.  We do not examine the subsequent quantization procedure, leaving this aspect for a future publication.
	\subsubsection{The Lagrangian and the equations of motion}	
	\par
	The Lagrangian density of macroscopic electromagnetism can be motivated as follows.  Firstly we have the familiar term associated with the electromagnetic field in vacuum,
	\begin{equation}
		\mathscr{L}_{\text{\tiny{F}}}=\frac{\epsilon_{0}}{2}\left[\boldsymbol{E}^{2}-c^{2}\boldsymbol{B}^{2}\right]\label{field_L}	
	\end{equation}
	where, \(\boldsymbol{E}=-\boldsymbol{\nabla}\phi-\dot{\boldsymbol{A}}\), and \(\boldsymbol{B}=\boldsymbol{\nabla}\boldsymbol{\times}\boldsymbol{A}\).
	\par
	All media are dispersive, and hence via the Kramers--Kronig relations exhibit significant loss at some frequencies.  When the medium does not depend explicitly on time, the Lagrangian must conserve energy, and this field energy lost from the dynamics of (\ref{field_L}) must be transferred into another system.  The response of the medium is assumed to have an arbitrary spatial dependence, so the additional system is proposed to be a reservoir of independent harmonic oscillators that exists at all points in space, and at each point contains every possible frequency of oscillator,
	\begin{equation}
		\mathscr{L}_{\text{\tiny{R}}}=\frac{1}{2}\int_{0}^{\infty}\left[\dot{\boldsymbol{X}}_{\omega}^{2}+\dot{\boldsymbol{Y}}_{\omega}^{2}-\omega^{2}\left(\boldsymbol{X}_{\omega}^{2}+\boldsymbol{Y}_{\omega}^{2}\right)\right]d\omega\label{resevoir}
	\end{equation}
	In the Lagrangian density associated with the reservoir, (\ref{resevoir}), two oscillators, \(\boldsymbol{X}_{\omega}\) and \(\boldsymbol{Y}_{\omega}\), are present at every frequency to account for the fact that we must distinguish between a loss of field energy through the electric interaction (the imaginary part of \(\boldsymbol{\epsilon}\)~\footnote{In general, loss is determined by the non--Hermitian part of the susceptibility.  However, the complex elements of a Hermitian susceptibility are related to time irreversibility, or spatial dispersion, effects which are not included in the theory of section \ref{summary-section}.}), and through the magnetic interaction (the imaginary part of \(\boldsymbol{\mu}\)).  
	\par
	Finally, the field must be coupled to the bath of oscillators in such a way that the classical macroscopic Maxwell equations give an extreme value for the action.  It is found that such a coupling is given by,
	\begin{equation}
		\mathscr{L}_{\text{\tiny{INT}}}=\boldsymbol{E}\boldsymbol{\cdot}\int_{0}^{\infty}\boldsymbol{\alpha}_{\text{\tiny{E\,E}}}(\omega)\boldsymbol{\cdot}\boldsymbol{X}_{\omega}d\omega+\boldsymbol{B}\boldsymbol{\cdot}\int_{0}^{\infty}\boldsymbol{\alpha}_{\text{\tiny{B\,B}}}(\omega)\boldsymbol{\cdot}\boldsymbol{Y}_{\omega}d\omega,\label{interaction}
	\end{equation}
	where the interaction with the material is determined via the coupling tensors, \(\boldsymbol{\alpha}_{\text{\tiny{E\,E}}}(\omega)\) and \(\boldsymbol{\alpha}_{\text{\tiny{B\,B}}}(\omega)\), which are assumed to be analytic functions of \(\omega\) in the upper half plane, and vanish in vacuum.  Note that all quantities appearing in (\ref{field_L}-\ref{interaction}) are implicitly functions of position, and all of the fields are also functions of time.
	\par
	The interaction, (\ref{interaction}) can be understood in terms of the local polarization, \(\boldsymbol{P}\), and magnetization, \(\boldsymbol{M}\), of the medium, 
	\begin{align}
		\boldsymbol{P}&=\int_{0}^{\infty}\boldsymbol{\alpha}_{\text{\tiny{E\,E}}}(\omega)\boldsymbol{\cdot}\boldsymbol{X}_{\omega}d\omega\nonumber\\
		\boldsymbol{M}&=\int_{0}^{\infty}\boldsymbol{\alpha}_{\text{\tiny{B\,B}}}(\omega)\boldsymbol{\cdot}\boldsymbol{Y}_{\omega}d\omega.\label{polarization-magnetization}
	\end{align}
	With this notation, (\ref{interaction}) takes the usual dipolar form, which would be expected from a local interaction with a neutral medium (see e.g.~\cite{horsley2006}).
	\par
	As usual, the action is equal to a four dimensional integral of the Lagrangian density, which in this case is given by \(\mathscr{L}=\)(\ref{field_L})\(+\)(\ref{resevoir})\(+\)(\ref{interaction}),
	\begin{equation}
		S[\phi,\boldsymbol{A},\boldsymbol{X}_{\omega},\boldsymbol{Y}_{\omega}]=\int\left[\mathcal{L}_{\text{\tiny{F}}}+\mathscr{L}_{\text{\tiny{R}}}+\mathscr{L}_{\text{\tiny{INT}}}\right]d^{4}x\label{em-action}
	\end{equation}
	The remarkable features of (\ref{em-action}) only become clear in the equations of motion, which are derived from finding an extremum of \(S\).
	\par
	Using the usual field equations~\cite{volume2}, each of the oscillators is found to evolve according to,
	\begin{align}
		\ddot{\boldsymbol{X}}_{\omega}&=-\omega^{2}\boldsymbol{X}_{\omega}+\boldsymbol{\alpha}_{\text{\tiny{E\,E}}}^{T}(\omega)\boldsymbol{\cdot}\boldsymbol{E}\nonumber\\
		\ddot{\boldsymbol{Y}}_{\omega}&=-\omega^{2}\boldsymbol{Y}_{\omega}+\boldsymbol{\alpha}_{\text{\tiny{B\,B}}}^{T}(\omega)\boldsymbol{\cdot}\boldsymbol{B},\label{oscillator-eqns}	
	\end{align}
	with the field obeying the usual Maxwell equations,
	\begin{align}
		\boldsymbol{\nabla}\boldsymbol{\cdot}\boldsymbol{D}&=0\nonumber\\
		\boldsymbol{\nabla}\boldsymbol{\times}\boldsymbol{H}&=\frac{\partial\boldsymbol{D}}{\partial t}	
	\end{align}
	where, \(\boldsymbol{D}=\epsilon_{0}\boldsymbol{E}+\boldsymbol{P}\), \& \(\boldsymbol{H}=\boldsymbol{B}/\mu_{0}-\boldsymbol{M}\), with \(\boldsymbol{P}\) and \(\boldsymbol{M}\) determined by the solution to (\ref{oscillator-eqns}), via (\ref{polarization-magnetization}).
	\par
	When the coupling, \(\boldsymbol{\alpha}_{\text{\tiny{E\,E}}}\), does not depend on time, the simplest way to solve (\ref{oscillator-eqns}) is in the frequency domain, where, for example, \(\boldsymbol{X}_{\omega}=\int\tilde{\boldsymbol{X}}_{\omega}(\Omega)\exp{(-i\Omega t)}d\Omega/2\pi\).  This leads to,
	\begin{multline}
		\tilde{\boldsymbol{X}}_{\omega}(\Omega)=\frac{\boldsymbol{\alpha}_{\text{\tiny{E\,E}}}^{T}(\omega)\boldsymbol{\cdot}\tilde{\boldsymbol{E}}(\Omega)}{(\omega+\Omega+i\eta)(\omega-\Omega-i\eta)}\\+2\pi\left[\boldsymbol{h}_{\text{\tiny{X}}\omega}\delta(\Omega-\omega)+\boldsymbol{h}_{\text{\tiny{X}}\omega}^{\star}\delta(\Omega+\omega)\right]\label{oscillator-solution-poles}
	\end{multline}
	where the \(\boldsymbol{h}_{\text{\tiny{X\,(Y)}}\omega}\) are solutions to the homogeneous equation.  A similar relation also holds for \(\tilde{\boldsymbol{Y}}_{\omega}\).  We understand the transformation of (\ref{oscillator-solution-poles}) into the time domain in the limit as, \(\eta\to0\), where the choice of poles has been constrained by , \(\tilde{\boldsymbol{X}}_{\omega}^{\star}(\Omega)=\tilde{\boldsymbol{X}}_{\omega}(-\Omega)\).  The choice of the sign of \(\eta\) corresponds to retarded versus advanced solutions.  As we shall see, in this case the choice also amounts to the physics of absorption versus gain.
	\par
	Transforming (\ref{oscillator-solution-poles}) into the time domain, along with the corresponding expression for \(\tilde{\boldsymbol{Y}}_{\omega}\), we insert these quantities into (\ref{polarization-magnetization}), and find the general evolution of \(\boldsymbol{P}\) and \(\boldsymbol{M}\) in terms of the field amplitudes and the \(\boldsymbol{h}_{\omega}\).  For the electric polarization, this gives,
	\begin{multline}
		\boldsymbol{P}=\text{P}\int_{0}^{\infty}\frac{d\Omega}{2\pi}\int_{0}^{\infty}d\omega\frac{\boldsymbol{\alpha}_{\text{\tiny{E\,E}}}(\omega)\boldsymbol{\cdot}\boldsymbol{\alpha}^{T}_{\text{\tiny{E\,E}}}(\omega)\boldsymbol{\cdot}\tilde{\boldsymbol{E}}(\Omega)e^{-i\Omega t}}{(\omega^{2}-\Omega^{2})}\\
		\pm i\int_{0}^{\infty}d\omega\frac{\boldsymbol{\alpha}_{\text{\tiny{E\,E}}}(\omega)\boldsymbol{\cdot}\boldsymbol{\alpha}^{T}_{\text{\tiny{E\,E}}}(\omega)\boldsymbol{\cdot}\tilde{\boldsymbol{E}}(\omega)e^{-i\omega t}}{4\omega}\\
		+\int_{0}^{\infty}d\omega\boldsymbol{\alpha}_{\text{\tiny{E\,E}}}(\omega)\boldsymbol{\cdot}\boldsymbol{h}_{\text{\tiny{X}}\omega}e^{-i\omega t}+\text{c.c.}\label{electric-polarization}
	\end{multline}
	where `\(\text{P}\)' denotes the Cauchy principal value of the integral, and the choice of sign comes from the choice of sign of \(\eta\) before the limit is taken.  If we consider (\ref{electric-polarization}) in the case of \(\text{sgn}(\eta)=+1\), then it becomes clear that if we write,
	\begin{equation}
		\boldsymbol{\alpha}_{\text{\tiny{E\,E\,(B\,B)}}}(\omega)\boldsymbol{\cdot}\boldsymbol{\alpha}^{T}_{\text{\tiny{E\,E\,(B\,B)}}}(\omega)=\frac{2\omega}{\pi}\Im{\left[\boldsymbol{\chi}_{\text{\tiny{E\,E\,(B\,B)}}}(\omega)\right]}\label{coupling-susceptibility}
	\end{equation}
	with, \(\boldsymbol{\chi}_{\text{\tiny{E\,E\,(B\,B)}}}(\omega)\) interpreted as the electric (magnetic) susceptibility~\footnote{Our definition of the magnetic susceptibility is \(\boldsymbol{H}(\omega)=\left(\mu_{0}^{-1}\boldsymbol{\mathbb{1}}_{3}-\boldsymbol{\chi}_{\text{\tiny{B}}}\right)\boldsymbol{\cdot}\boldsymbol{B}(\omega)\).}, then the Kramers--Kronig relations~\cite{volume8} are automatically satisfied,
	\begin{equation}
		\Re\left[\boldsymbol{\chi}_{\text{\tiny{E\,E\,(B\,B)}}}(\omega)\right]=\frac{2}{\pi}\text{P}\int_{0}^{\infty}\frac{\Omega\Im\left[\boldsymbol{\chi}_{\text{\tiny{E\,E\,(B\,B)}}}(\Omega)\right]}{\Omega^{2}-\omega^{2}}d\Omega,\label{kramers-kronig}	
	\end{equation}
	and that the electric polarization in (\ref{electric-polarization}) can be written in the usual form,
	\begin{equation}
		\boldsymbol{P}=\boldsymbol{P}_{0}+\int_{-\infty}^{\infty}\frac{d\omega}{2\pi}\boldsymbol{\chi}_{\text{\tiny{E\,E}}}(\omega)\boldsymbol{\cdot}\tilde{\boldsymbol{E}}(\omega)e^{-i\omega t},\label{final-polarization}	
	\end{equation}
	where, \(\boldsymbol{P}_{0}=\int_{0}^{\infty}\boldsymbol{\alpha}_{\text{\tiny{E\,E}}}(\omega)\boldsymbol{\cdot}\boldsymbol{h}_{\text{\tiny{X}}\omega}\exp{(-i\omega t)}d\omega+\text{c.c.}\) is the `undriven' part of the polarization, that does not depend on the electromagnetic field.  Similar relations hold for the magnetization, \(\boldsymbol{M}\), with the substitutions \(E\to B\), and \(X\to Y\).  It is striking that, with the identification given in (\ref{coupling-susceptibility}), the classical equations of motion arising from (\ref{em-action}) can represent electromagnetism within \emph{any} material that may be characterised by linear electric and magnetic susceptibilities.  
	\par
	The fact that the theory can be written in terms of a Lagrangian containing only first order time derivatives, from which we can derive a Hamiltonian, enables it to be quantized canonically; with Poisson or Dirac brackets becoming commutators.  In quantizing the field, \(\boldsymbol{P}_{0}\) and \(\boldsymbol{M}_{0}\) are present by necessity: the classical amplitude, \(\boldsymbol{h}_{\omega}\) becomes an operator, related to the fluctuating `noise currents' already found to be necessary within the phenomenological theory (e.g.~\cite{dung1998}).  Furthermore, as the Kramers--Kronig relations, (\ref{kramers-kronig}), arise from the dynamics of the reservoir of oscillators, it should be possible to apply this technique to quantize any theory of linear response.
	\subsubsection{Restrictions on the constitutive relations, and the origin of loss versus gain}
	\par
	If we define the coupling to the reservoir via (\ref{coupling-susceptibility}), then the susceptibility tensors are only consistent with the Lagrangian if they are symmetric; a requirement usually arising from statistical physics~\cite{volume5,volume8}, but automatically fulfilled here.  Due to this symmetry of the susceptibility tensor there is also a well defined procedure for determining \(\boldsymbol{\alpha}_{\text{\tiny{E\,E\,(B\,B)}}}\) from \(\Im[\boldsymbol{\chi}_{\text{\tiny{E\,E\,(B\,B)}}}]\), although the result will not be unique~\footnote{As \(\Im[\boldsymbol{\chi}_{\text{\tiny{E\,E\,(B\,B)}}}]\) is symmetric and real, it is a normal matrix, which can be diagonalised with an orthogonal matrix, \(\boldsymbol{O}\): \(\Im[\boldsymbol{\chi}_{\text{\tiny{E\,E\,(B\,B)}}}]=\boldsymbol{O}^{T}\boldsymbol{D}\boldsymbol{O}\), where \(\boldsymbol{D}=\text{diag}(\lambda_{1},\lambda_{2},\lambda_{3})\).  It is therefore possible to write down an expression for the coupling, \(\boldsymbol{\alpha}_{\text{E\,E\,(B\,B)}}=\boldsymbol{O}^{T}\boldsymbol{D}^{1/2}\), where \(\boldsymbol{D}^{1/2}=\text{diag}(\pm\sqrt{\lambda}_{1},\pm\sqrt{\lambda}_{2},\pm\sqrt{\lambda}_{3})\).  This expression is not unique, due to the eight possible choices of sign, and is only real if \(\Im[\boldsymbol{\chi}_{\text{\tiny{E\,E\,(B\,B)}}}]\) is positive definite (\(\lambda_{i}>0\)).}.  The procedure also only produces a real \(\boldsymbol{\alpha}_{\text{\tiny{E\,E\,(B\,B)}}}\) if \(\Im[\boldsymbol{\chi}_{\text{\tiny{E\,E\,(B\,B)}}}]\) is positive definite.
	\par
	The lack of uniqueness of the coupling tensors plays no role in the classical theory, as \(\boldsymbol{\alpha}_{\text{\tiny{E\,E\,(B\,B)}}}\) only appears linearly in \(\boldsymbol{P}_{0}\), which may be given any form through a suitable choice of \(\boldsymbol{h}_{\omega}\).  However, when the system is quantized the equivalent of the `noise current' operator is given in terms of the operator versions of \(\boldsymbol{P}_{0}\) and \(\boldsymbol{M}_{0}\) (with the \(\boldsymbol{h}_{\omega}\) now becoming operators with certain commutation relations)~\cite{philbin2010}, and therefore the choice of \(\boldsymbol{\alpha}_{\text{\tiny{E\,E\,(B\,B)}}}\) will make a difference to the vacuum fluctuation of the polarization and magnetization of the medium.  It is not clear whether this difference has any observable consequences, although we do not address this problem here.
	\par
	Another interesting feature is that the sign of the imaginary part of \(\boldsymbol{\chi}_{\text{\tiny{E\,E\,(B\,B)}}}\) in (\ref{electric-polarization}) is determined by a choice made in the dynamics of the reservoir.  Therefore the distinction between loss and gain in this theory does not appear at the level of the Lagrangian, but in the boundary conditions imposed on the equations of motion of the reservoir.
%
%
	\section{An extension to other reservoir--field coupling terms\label{exploration-section}}
	\par
	As stated in the introduction, there are several reasons to investigate an extension of the Lagrangian theory of section \ref{summary-section} to magnetoelectric media.  There is a history of controversy regarding not only bounds on the magnetoelectric susceptibilities, \(\boldsymbol{\chi}_{\text{\tiny{E\,B}}}\) \& \(\boldsymbol{\chi}_{\text{\tiny{B\,E}}}\), but also the kinds of magnetoelectric coupling that are possible in principle~\cite{lakhtakia1994b,lakhtakia1994c,raab1995,weiglhofer1998}.  Moreover, it is not clear to what extent moving media and magnetoelectrics are equivalent, something which is important in the discussion of a `frictional' component to the Casimir force~\cite{philbin2009,leonhardt2010,pendry2010}.  Finding a Lagrangian that describes these materials and can be quantized may help clarify these issues.
	\par
	If there are macroscopic parameters that prove inconsistent with a Lagrangian (or more precisely, a Hamiltonian), the electromagnetic field cannot be quantized within such media in an obvious way, and we take this as a restriction on the material parameters.
	\par
	What additional coupling terms could be added into (\ref{interaction})?  Lagrangians containing time derivatives of the fields higher than the first (which cannot be removed with a gauge transformation) do not have well defined canonical momenta, and consequently we cannot derive a Hamiltonian.  If we also consider the polarization and magnetization to only depend upon the local values of the oscillators, \(\boldsymbol{X}_{\omega}\) \& \(\boldsymbol{Y}_{\omega}\), then we have the following generalization of (\ref{polarization-magnetization}),
	\begin{multline}
		\boldsymbol{P}=\int_{0}^{\infty}\Bigg[\left(\boldsymbol{\alpha}_{\text{\tiny{E\,E}}}+\boldsymbol{\beta}_{\text{\tiny{E\,E}}}\frac{\partial}{\partial t}\right)\boldsymbol{\cdot}\boldsymbol{X}_{\omega}\\
		+\left(\boldsymbol{\alpha}_{\text{\tiny{E\,B}}}+\boldsymbol{\beta}_{\text{\tiny{E\,B}}}\frac{\partial}{\partial t}\right)\boldsymbol{\cdot}\boldsymbol{Y}_{\omega}\Bigg]d\omega\label{all-couplings-1}
	\end{multline}
	and,
	\begin{multline}
		\boldsymbol{M}=\int_{0}^{\infty}\Bigg[\left(\boldsymbol{\alpha}_{\text{\tiny{B\,B}}}+\boldsymbol{\beta}_{\text{\tiny{B\,B}}}\frac{\partial}{\partial t}\right)\boldsymbol{\cdot}\boldsymbol{Y}_{\omega}\\
		+\left(\boldsymbol{\alpha}_{\text{\tiny{B\,E}}}+\boldsymbol{\beta}_{\text{\tiny{B\,E}}}\frac{\partial}{\partial t}\right)\boldsymbol{\cdot}\boldsymbol{X}_{\omega}\Bigg]d\omega\label{all-couplings-2}	
	\end{multline}
	Table \ref{symmetry-table} shows the properties of the coupling tensors that arise from assuming that the value of the action is separately invariant under an active time reversal (\(t\to-t\)) and space inversion (\(\boldsymbol{x}\to-\boldsymbol{x}\)) of the fields; e.g. \(\boldsymbol{X}_{\omega}(\boldsymbol{x},t)\to\boldsymbol{X}_{\omega}(\boldsymbol{x},-t)\) \& \(\boldsymbol{X}_{\omega}(\boldsymbol{x},t)\to\boldsymbol{X}_{\omega}(-\boldsymbol{x},t)\).  In obtaining table \ref{symmetry-table}, we have assumed that the \(\boldsymbol{X}_{\omega}\) \& \(\boldsymbol{Y}_{\omega}\) oscillator amplitudes obey the same symmetry properties as the electric and magnetic fields, respectively.
	\par
	If a coupling tensor is non-zero and changes sign under time reversal, then the medium possesses an intrinsic time irreversibility (e.g. due to an external magnetic field, or motion).  Meanwhile a change of sign under spatial inversion shows that the medium possesses a certain handedness (e.g. chiral media).
	\par
	Table \ref{symmetry-table} therefore shows that magnetoelectrics violating spatial inversion symmetry, while exhibiting time--reversibility, such as those constructed from chiral inclusions, must be described by the coupling terms \(\boldsymbol{\beta}_{\text{\tiny{E\,B}}}\) and \(\boldsymbol{\beta}_{\text{\tiny{B\,E}}}\).  On the other hand, inversion symmetric, time irreversible media, such as a dielectric in an external magnetic field (a Faraday medium) must be described by \(\boldsymbol{\beta}_{\text{\tiny{E\,E}}}\) \& \(\boldsymbol{\beta}_{\text{\tiny{B\,B}}}\). Finally, time--irreversible magnetoelectrics that also violate spatial inversion symmetry, such as moving media, or Tellegen media~\cite{tellegen1948} must arise from \(\boldsymbol{\alpha}_{\text{\tiny{E\,B}}}\) and \(\boldsymbol{\alpha}_{\text{\tiny{B\,E}}}\).
	\par
	\begin{table}
	\begin{tabular}[c]{l|c|c}
		&\(\boldsymbol{x}\to-\boldsymbol{x}\)&\(t\to-t\)\\
		\hline
		\(\boldsymbol{\alpha}_{\text{\tiny{E\,E}}}\)&\(+\)&\(+\)\\
		\(\boldsymbol{\alpha}_{\text{\tiny{E\,B}}}\)&\(-\)&\(-\)\\
		\(\boldsymbol{\alpha}_{\text{\tiny{B\,B}}}\)&\(+\)&\(+\)\\
		\(\boldsymbol{\alpha}_{\text{\tiny{B\,E}}}\)&\(-\)&\(-\)\\	
		\hline
		\(\boldsymbol{\beta}_{\text{\tiny{E\,E}}}\)&\(+\)&\(-\)\\
		\(\boldsymbol{\beta}_{\text{\tiny{E\,B}}}\)&\(-\)&\(+\)\\	
		\(\boldsymbol{\beta}_{\text{\tiny{B\,B}}}\)&\(+\)&\(-\)\\
		\(\boldsymbol{\beta}_{\text{\tiny{B\,E}}}\)&\(-\)&\(+\)\\
	\end{tabular}
	\caption{Symmetry of coupling tensors under spatial inversion and time reversal.  A plus sign indicates that the coupling tensor does not change under the inversion operation, and a minus sign indicates that the tensor is multiplied by \(-1\).\label{symmetry-table}}
	\end{table}
	We now examine the macroscopic theory that results from using the interaction Lagrangian with the polarization and magnetization, (\ref{all-couplings-1}--\ref{all-couplings-2}).
	\subsection{The evolution of the oscillator amplitudes}	
	With the interaction Lagrangian defined by (\ref{all-couplings-1}--\ref{all-couplings-2}) as in (\ref{interaction}), the equations of motion for each of the oscillators in the reservoir are now,
	\begin{multline}
		\ddot{\boldsymbol{X}}_{\omega}=-\omega^{2}\boldsymbol{X}_{\omega}+\left(\boldsymbol{\alpha}_{\text{\tiny{E\,E}}}^{T}-\boldsymbol{\beta}_{\text{\tiny{E\,E}}}^{T}\frac{\partial}{\partial t}\right)\boldsymbol{\cdot}\boldsymbol{E}\\
		+\left(\boldsymbol{\alpha}^{T}_{\text{\tiny{B\,E}}}-\boldsymbol{\beta}^{T}_{\text{\tiny{B\,E}}}\frac{\partial}{\partial t}\right)\boldsymbol{\cdot}\boldsymbol{B}
	\end{multline}
	and,
	\begin{multline}
		\ddot{\boldsymbol{Y}}_{\omega}=-\omega^{2}\boldsymbol{Y}_{\omega}+\left(\boldsymbol{\alpha}_{\text{\tiny{B\,B}}}^{T}-\boldsymbol{\beta}_{\text{\tiny{B\,B}}}^{T}\frac{\partial}{\partial t}\right)\boldsymbol{\cdot}\boldsymbol{B}\\
		+\left(\boldsymbol{\alpha}^{T}_{\text{\tiny{E\,B}}}-\boldsymbol{\beta}^{T}_{\text{\tiny{E\,B}}}\frac{\partial}{\partial t}\right)\boldsymbol{\cdot}\boldsymbol{E}
	\end{multline}
	In Fourier space, we therefore obtain,
	\begin{align}
		\tilde{\boldsymbol{X}}_{\omega}(\Omega)&=\frac{\left(\boldsymbol{\alpha}_{\text{\tiny{E\,E}}}^{T}+i\Omega\boldsymbol{\beta}_{\text{\tiny{E\,E}}}^{T}\right)\boldsymbol{\cdot}\tilde{\boldsymbol{E}}+\left(\boldsymbol{\alpha}^{T}_{\text{\tiny{B\,E}}}+i\Omega\boldsymbol{\beta}^{T}_{\text{\tiny{B\,E}}}\right)\boldsymbol{\cdot}\tilde{\boldsymbol{B}}}{\left(\omega-\Omega-i\eta\right)\left(\omega+\Omega+i\eta\right)}\nonumber\\
		&+\dots\nonumber\\
		\tilde{\boldsymbol{Y}}_{\omega}(\Omega)&=\frac{\left(\boldsymbol{\alpha}_{\text{\tiny{B\,B}}}^{T}+i\Omega\boldsymbol{\beta}_{\text{\tiny{B\,B}}}^{T}\right)\boldsymbol{\cdot}\tilde{\boldsymbol{B}}+\left(\boldsymbol{\alpha}^{T}_{\text{\tiny{E\,B}}}+i\Omega\boldsymbol{\beta}^{T}_{\text{\tiny{E\,B}}}\right)\boldsymbol{\cdot}\tilde{\boldsymbol{E}}}{\left(\omega-\Omega-i\eta\right)\left(\omega+\Omega+i\eta\right)}\nonumber\\
		&+\dots\label{fourier-solution}
	\end{align}
	with the poles dealt with as in section~\ref{summary-section}, and the homogeneous parts of the solution omitted (c.f. (\ref{oscillator-solution-poles})).  Note that from now on we consider the case of media with loss rather than gain.
	\par
	Transforming (\ref{fourier-solution}) into the time domain gives us the final expressions for the evolution of the reservoir.  The \(\boldsymbol{X}_{\omega}\) oscillator obeys,
	\begin{widetext}
	\begin{multline}
		\boldsymbol{X}_{\omega}=\text{P}\int_{0}^{\infty}\frac{d\Omega}{2\pi}\left[\frac{\left(\boldsymbol{\alpha}_{\text{\tiny{E\,E}}}^{T}+i\Omega\boldsymbol{\beta}_{\text{\tiny{E\,E}}}^{T}\right)\boldsymbol{\cdot}\tilde{\boldsymbol{E}}+\left(\boldsymbol{\alpha}_{\text{\tiny{B\,E}}}^{T}+i\Omega\boldsymbol{\beta}_{\text{\tiny{B\,E}}}^{T}\right)\boldsymbol{\cdot}\tilde{\boldsymbol{B}}}{(\omega^{2}-\Omega^{2})}\right]e^{-i\Omega t}\\
		+\frac{i}{4\omega}\left[\left(\boldsymbol{\alpha}^{T}_{\text{\tiny{E\,E}}}+i\omega\boldsymbol{\beta}^{T}_{\text{\tiny{E\,E}}}\right)\boldsymbol{\cdot}\tilde{\boldsymbol{E}}+\left(\boldsymbol{\alpha}^{T}_{\text{\tiny{B\,E}}}+i\omega\boldsymbol{\beta}^{T}_{\text{\tiny{B\,E}}}\right)\boldsymbol{\cdot}\tilde{\boldsymbol{B}}\right]e^{-i\omega t}+\boldsymbol{h}_{\text{\tiny{X}}\omega}e^{-i\omega t}+\text{c.c.},\label{e-osc}
	\end{multline}
	and the \(\boldsymbol{Y}_{\omega}\),
	\begin{multline}
		\boldsymbol{Y}_{\omega}=\text{P}\int_{0}^{\infty}\frac{d\Omega}{2\pi}\left[\frac{\left(\boldsymbol{\alpha}_{\text{\tiny{B\,B}}}^{T}+i\Omega\boldsymbol{\beta}_{\text{\tiny{B\,B}}}^{T}\right)\boldsymbol{\cdot}\tilde{\boldsymbol{B}}+\left(\boldsymbol{\alpha}_{\text{\tiny{E\,B}}}^{T}+i\Omega\boldsymbol{\beta}_{\text{\tiny{E\,B}}}^{T}\right)\boldsymbol{\cdot}\tilde{\boldsymbol{E}}}{(\omega^{2}-\Omega^{2})}\right]e^{-i\Omega t}\\
		+\frac{i}{4\omega}\left[\left(\boldsymbol{\alpha}^{T}_{\text{\tiny{B\,B}}}+i\omega\boldsymbol{\beta}^{T}_{\text{\tiny{B\,B}}}\right)\boldsymbol{\cdot}\tilde{\boldsymbol{B}}+\left(\boldsymbol{\alpha}^{T}_{\text{\tiny{E\,B}}}+i\omega\boldsymbol{\beta}^{T}_{\text{\tiny{E\,B}}}\right)\boldsymbol{\cdot}\tilde{\boldsymbol{E}}\right]e^{-i\omega t}+\boldsymbol{h}_{\text{\tiny{Y}}\omega}e^{-i\omega t}+\text{c.c.}.\label{b-osc}
	\end{multline}
	\end{widetext}
	Equations (\ref{e-osc}--\ref{b-osc}) now determine the evolution of \(\boldsymbol{P}\) \& \(\boldsymbol{M}\) in terms of the field amplitudes.
%
%
	\subsection{The polarization and magnetization in terms of the field amplitudes\label{P-M-subsection}}
	Inserting (\ref{e-osc}--\ref{b-osc}) into (\ref{all-couplings-1}--\ref{all-couplings-2}) gives us a polarization and magnetization of the form,
	\begin{align}
		\boldsymbol{P}=\boldsymbol{P}_{0}+\int_{-\infty}^{\infty}\left[\boldsymbol{\chi}_{\text{\tiny{E\,E}}}(\omega)\boldsymbol{\cdot}\tilde{\boldsymbol{E}}+\boldsymbol{\chi}_{\text{\tiny{E\,B}}}(\omega)\boldsymbol{\cdot}\tilde{\boldsymbol{B}}\right]e^{-i\omega t}\frac{d\omega}{2\pi}\nonumber\\
		\boldsymbol{M}=\boldsymbol{M}_{0}+\int_{-\infty}^{\infty}\left[\boldsymbol{\chi}_{\text{\tiny{B\,B}}}(\omega)\boldsymbol{\cdot}\tilde{\boldsymbol{B}}+\boldsymbol{\chi}_{\text{\tiny{B\,E}}}(\omega)\boldsymbol{\cdot}\tilde{\boldsymbol{E}}\right]e^{-i\omega t}\frac{d\omega}{2\pi}\label{magnetoelectric-constitutive}	
	\end{align}
	where the electric--electric and magnetic--magnetic susceptibilities are given by,
	\begin{align}
		\boldsymbol{\chi}_{\text{\tiny{E\,E}}}(\omega)=&\text{P}\int_{0}^{\infty}d\Omega\frac{\boldsymbol{\lambda}_{\text{\tiny{E\,E}}}\cdot\boldsymbol{\lambda}_{\text{\tiny{E\,E}}}^{\dagger}(\omega,\Omega)+\boldsymbol{\lambda}_{\text{\tiny{E\,B}}}\cdot\boldsymbol{\lambda}^{\dagger}_{\text{\tiny{E\,B}}}(\omega,\Omega)}{(\Omega^{2}-\omega^{2})}\nonumber\\
		&+\frac{i\pi}{2\omega}[\boldsymbol{\lambda}_{\text{\tiny{E\,E}}}\cdot\boldsymbol{\lambda}_{\text{\tiny{E\,E}}}^{\dagger}(\omega,\omega)+\boldsymbol{\lambda}_{\text{\tiny{E\,B}}}\cdot\boldsymbol{\lambda}^{\dagger}_{\text{\tiny{E\,B}}}(\omega,\omega)]\nonumber\\
		\boldsymbol{\chi}_{\text{\tiny{B\,B}}}(\omega)=&\text{P}\int_{0}^{\infty}d\Omega\frac{\boldsymbol{\lambda}_{\text{\tiny{B\,B}}}\cdot\boldsymbol{\lambda}_{\text{\tiny{B\,B}}}^{\dagger}(\omega,\Omega)+\boldsymbol{\lambda}_{\text{\tiny{B\,E}}}\cdot\boldsymbol{\lambda}^{\dagger}_{\text{\tiny{B\,E}}}(\omega,\Omega)}{(\Omega^{2}-\omega^{2})}\nonumber\\
		&+\frac{i\pi}{2\omega}[\boldsymbol{\lambda}_{\text{\tiny{B\,B}}}\cdot\boldsymbol{\lambda}_{\text{\tiny{B\,B}}}^{\dagger}(\omega,\omega)+\boldsymbol{\lambda}_{\text{\tiny{B\,E}}}\cdot\boldsymbol{\lambda}^{\dagger}_{\text{\tiny{B\,E}}}(\omega,\omega)]\label{eebb-sus}
	\end{align}	
	and the magnetoelectric susceptibilities are,
	\begin{align}
	\boldsymbol{\chi}_{\text{\tiny{E\,B}}}(\omega)=&\text{P}\int_{0}^{\infty}d\Omega\frac{\boldsymbol{\lambda}_{\text{\tiny{E\,E}}}\cdot\boldsymbol{\lambda}_{\text{\tiny{B\,E}}}^{\dagger}(\omega,\Omega)+\boldsymbol{\lambda}_{\text{\tiny{E\,B}}}\cdot\boldsymbol{\lambda}^{\dagger}_{\text{\tiny{B\,B}}}(\omega,\Omega)}{(\Omega^{2}-\omega^{2})}\nonumber\\
		&+\frac{i\pi}{2\omega}[\boldsymbol{\lambda}_{\text{\tiny{E\,E}}}\cdot\boldsymbol{\lambda}_{\text{\tiny{B\,E}}}^{\dagger}(\omega,\omega)+\boldsymbol{\lambda}_{\text{\tiny{E\,B}}}\cdot\boldsymbol{\lambda}^{\dagger}_{\text{\tiny{B\,B}}}(\omega,\omega)]\nonumber\\
		\boldsymbol{\chi}_{\text{\tiny{B\,E}}}(\omega)=&\text{P}\int_{0}^{\infty}d\Omega\frac{\boldsymbol{\lambda}_{\text{\tiny{B\,B}}}\cdot\boldsymbol{\lambda}_{\text{\tiny{E\,B}}}^{\dagger}(\omega,\Omega)+\boldsymbol{\lambda}_{\text{\tiny{B\,E}}}\cdot\boldsymbol{\lambda}^{\dagger}_{\text{\tiny{E\,E}}}(\omega,\Omega)}{(\Omega^{2}-\omega^{2})}\nonumber\\
		&+\frac{i\pi}{2\omega}[\boldsymbol{\lambda}_{\text{\tiny{B\,B}}}\cdot\boldsymbol{\lambda}_{\text{\tiny{E\,B}}}^{\dagger}(\omega,\omega)+\boldsymbol{\lambda}_{\text{\tiny{B\,E}}}\cdot\boldsymbol{\lambda}^{\dagger}_{\text{\tiny{E\,E}}}(\omega,\omega)]\label{ebbe-sus}
	\end{align}
	where,
	\begin{align}
		\boldsymbol{\lambda}_{\text{\tiny{E\,E}}}(\omega,\Omega)=\boldsymbol{\alpha}_{\text{\tiny{E\,E}}}(\Omega)-i\omega\boldsymbol{\beta}_{\text{\tiny{E\,E}}}(\Omega)\nonumber\\
		\boldsymbol{\lambda}_{\text{\tiny{B\,B}}}(\omega,\Omega)=\boldsymbol{\alpha}_{\text{\tiny{B\,B}}}(\Omega)-i\omega\boldsymbol{\beta}_{\text{\tiny{B\,B}}}(\Omega)\nonumber\\
		\boldsymbol{\lambda}_{\text{\tiny{E\,B}}}(\omega,\Omega)=\boldsymbol{\alpha}_{\text{\tiny{E\,B}}}(\Omega)-i\omega\boldsymbol{\beta}_{\text{\tiny{E\,B}}}(\Omega)\nonumber\\
		\boldsymbol{\lambda}_{\text{\tiny{B\,E}}}(\omega,\Omega)=\boldsymbol{\alpha}_{\text{\tiny{B\,E}}}(\Omega)-i\omega\boldsymbol{\beta}_{\text{\tiny{B\,E}}}(\Omega)\label{lambda-tensors}
	\end{align}
	The physical interpretation of the dependence of (\ref{lambda-tensors}) on two frequencies, \(\omega\) and \(\Omega\), is that in general the medium responds to the field amplitudes and the \emph{rate of change} of the field amplitudes.  Therefore, in the integrals within (\ref{eebb-sus}--\ref{ebbe-sus}) we have factors of \(\omega\) that arise from a linear response to a change in the field (e.g. a polarization due to the EMF from a changing magnetic field), and factors of \(\Omega\) that relate to the dispersion due to the finite response time of the medium.  
\par	
	With the results (\ref{eebb-sus}--\ref{ebbe-sus}), we can see that the susceptibilities are divided into parts related to dissipation, and lossless response, the connection between the two being an interesting generalization of the Kramers--Kronig relation (\ref{kramers-kronig}).
\par	
	For the electric--electric and magnetic--magnetic susceptibilities, the dissipation is determined by the anti--Hermitian part of the susceptibility (c.f.~\cite{volume8}), while the magnetoelectric dissipation is determined by \(\boldsymbol{\chi}_{\text{\tiny{E\,B}}}-\boldsymbol{\chi}_{\text{\tiny{B\,E}}}^{\dagger}\) (as can be verified through substituting the magnetoelectric constitutive relations into \textsection{80} of~\cite{volume8}).  It is interesting that these two susceptibilities are automatically related in such a way that a non--zero \(\boldsymbol{\chi}_{\text{\tiny{E\,B}}}\) necessitates a non--zero \(\boldsymbol{\chi}_{\text{\tiny{B\,E}}}\).

	 
	\par
	The undriven parts of the polarization and magnetization in (\ref{magnetoelectric-constitutive}) involve a coupling between the two reservoirs,
	\begin{multline}
		\boldsymbol{P}_{0}=\int_{0}^{\infty}\big[\boldsymbol{\lambda}_{\text{\tiny{E\,E}}}(\omega,\omega)\boldsymbol{\cdot}\boldsymbol{h}_{\text{\tiny{X}}\omega}+\boldsymbol{\lambda}_{\text{\tiny{E\,B}}}(\omega,\omega)\boldsymbol{\cdot}\boldsymbol{h}_{\text{\tiny{Y}}\omega}\big]e^{-i\omega t}d\omega\\
		+\text{c.c.}
	\end{multline}
	and,
	\begin{multline}
		\boldsymbol{M}_{0}=\int_{0}^{\infty}\big[\boldsymbol{\lambda}_{\text{\tiny{B\,B}}}(\omega,\omega)\boldsymbol{\cdot}\boldsymbol{h}_{\text{\tiny{X}}\omega}+\boldsymbol{\lambda}_{\text{\tiny{B\,E}}}(\omega,\omega)\boldsymbol{\cdot}\boldsymbol{h}_{\text{\tiny{Y}}\omega}\big]e^{-i\omega t}d\omega\\
		+\text{c.c.}
	\end{multline}
	Having now established a general form for \(\boldsymbol{P}\) \& \(\boldsymbol{M}\), we develop the physical interpretation of the various coupling tensors within the Lagrangian.
%
%
	\subsection{Physical interpretation}
	The constitutive relations (\ref{magnetoelectric-constitutive}) are, as anticipated, those of a magnetoelectric.  However, this is a very general form, where the material neither obeys time reversal, nor spatial inversion symmetry.  We now consider three special cases of (\ref{eebb-sus}--\ref{ebbe-sus}); time reversible, odd parity media; time irreversible, even parity media; and media that exhibit a static magnetoelectric response.
%
%
	\subsubsection{Time reversible, odd parity media\label{chiral-case}}
	\par
	If we demand that the medium be time--reversible, then, from table, \ref{symmetry-table}, \(\boldsymbol{\alpha}_{\text{\tiny{E\,B}}}=\boldsymbol{\alpha}_{\text{\tiny{B\,E}}}=\boldsymbol{\beta}_{\text{\tiny{E\,E}}}=\boldsymbol{\beta}_{\text{\tiny{B\,B}}}=\boldsymbol{0}\), and we obtain the frequency domain constitutive relations,
	\begin{align*}
		\tilde{\boldsymbol{D}}&=\boldsymbol{\epsilon}\boldsymbol{\cdot}\tilde{\boldsymbol{E}}+i\omega\boldsymbol{\kappa}\boldsymbol{\cdot}\tilde{\boldsymbol{B}}\\
		\tilde{\boldsymbol{H}}&=\boldsymbol{\mu}^{-1}\boldsymbol{\cdot}\tilde{\boldsymbol{B}}+i\omega\boldsymbol{\kappa}^{T}\boldsymbol{\cdot}\tilde{\boldsymbol{E}}
	\end{align*}
	where, \(\boldsymbol{\epsilon}=\epsilon_{0}\mathbb{1}_{3}+\boldsymbol{\chi}_{\text{\tiny{E\,E}}}\), \(\boldsymbol{\mu}^{-1}=\mu_{0}^{-1}\mathbb{1}_{3}-\boldsymbol{\chi}_{\text{\tiny{B\,B}}}\), and \(2\omega\Im(\boldsymbol{\kappa})=\boldsymbol{\alpha}_{\text{\tiny{E\,E}}}\boldsymbol{\cdot}\boldsymbol{\beta}^{T}_{\text{\tiny{B\,E}}}-\boldsymbol{\beta}_{\text{\tiny{E\,B}}}\boldsymbol{\cdot}\boldsymbol{\alpha}_{\text{\tiny{B\,B}}}^{T}\), and the real and imaginary parts of \(\boldsymbol{\kappa}\) are connected by (\ref{kramers-kronig}).  Consistent with the time reversibility of this situation, the permeability and permittivity are symmetric tensors.
	\par
	These are the constitutive relations of an anisotropic \emph{chiral} medium in Boys--Post form~\cite{lakhtakia1994}.  Notice the necessity of having \(\boldsymbol{\kappa}^{T}\) appearing in \(\tilde{\boldsymbol{H}}\), versus \(\boldsymbol{\kappa}\) appearing in \(\tilde{\boldsymbol{D}}\).  No other choices of \(\boldsymbol{\kappa}\) appear to be consistent with the theory.  The prefactor of \(\omega\) in the magnetoelectric coupling is crucial, as it has the consequence that in the static limit (\(\omega\to0\)), unless \(\boldsymbol{\kappa}\) diverges, the medium no longer has a magnetoelectric response.
%
%
	\subsubsection{Time irreversible, even parity media}
	\par
	Applying table \ref{symmetry-table} in this case means that \(\boldsymbol{\alpha}_{\text{\tiny{E\,B}}}=\boldsymbol{\alpha}_{\text{\tiny{B\,E}}}=\boldsymbol{\beta}_{\text{\tiny{E\,B}}}=\boldsymbol{\beta}_{\text{\tiny{B\,E}}}=\boldsymbol{0}\), and the magnetoelectric susceptibilities, (\ref{ebbe-sus}) vanish.  We are then left with,
	\begin{align}
		\tilde{\boldsymbol{D}}=\boldsymbol{\epsilon}\boldsymbol{\cdot}\tilde{\boldsymbol{E}}\nonumber\\
		\tilde{\boldsymbol{H}}=\boldsymbol{\mu}^{-1}\boldsymbol{\cdot}\tilde{\boldsymbol{B}}
	\end{align}
	where \(\boldsymbol{\epsilon}=\epsilon_{0}\mathbb{1}_{3}+\boldsymbol{\chi}_{\text{\tiny{E\,E}}}\), and \(\boldsymbol{\mu}^{-1}=\mu_{0}\mathbb{1}_{3}-\boldsymbol{\chi}_{\text{\tiny{B\,B}}}\).
	\par
	The consequence of having broken the time reversibility of the medium is that at frequencies where the loss is negligible, \(\boldsymbol{\epsilon}=\boldsymbol{\epsilon}^{\dagger}\) and \(\boldsymbol{\mu}=\boldsymbol{\mu}^{\dagger}\).  The fact that these tensors become Hermitian is consistent with the generalized principle of the symmetry of kinetic coefficients in the case when time reversal symmetry is broken, and the medium is without loss~\cite{volume5,volume8}.  Again, results from statistical physics emerge from this Lagrangian description.  As \(\omega\to0\), \(\boldsymbol{\epsilon}\) and \(\boldsymbol{\mu}\) become symmetric tensors.
	\par
	One physical example of a medium which would be described by a \(\boldsymbol{\beta}_{\text{\tiny{E\,E}}}\) term would be an ordinary dielectric in an external magnetic field.  For frequencies where such a medium is without loss, \(\boldsymbol{\epsilon}\) is a Hermitian tensor~\cite{volume8}.
%
%
	\subsubsection{Media exhibiting a static magnetoelectric response\label{tellegen-subsection}}
	\par
	In the limit when the electromagnetic field becomes static (\(\omega\to0\)), all of the quantities in (\ref{lambda-tensors}) become real, and only the \(\boldsymbol{\alpha}\) coupling tensors play a role.  The reason for considering this limit is that here we isolate the terms in the Lagrangian that describe the physics of media in motion, and distinguish Tellegen from chiral media (something that has caused controversy in the past~\cite{lakhtakia1994c,weiglhofer1998}).  The static field constitutive relations are,
	\begin{align}
		\tilde{\boldsymbol{D}}(\omega\to0)&=\boldsymbol{\epsilon}\boldsymbol{\cdot}\tilde{\boldsymbol{E}}+\boldsymbol{\chi}_{\text{\tiny{E\,B}}}\boldsymbol{\cdot}\boldsymbol{B}\nonumber\\
		\tilde{\boldsymbol{H}}(\omega\to0)&=\boldsymbol{\mu}^{-1}\boldsymbol{\cdot}\tilde{\boldsymbol{B}}-\boldsymbol{\chi}^{T}_{\text{\tiny{E\,B}}}\boldsymbol{\cdot}\boldsymbol{E}\label{tellegen-cons}
	\end{align}
	Observe that, if a medium is to have a static magnetoelectric response, then this must be characterised with the coupling terms containing the \(\boldsymbol{\alpha}_{\text{\tiny{E\,B}}}\) and \(\boldsymbol{\alpha}_{\text{\tiny{B\,E}}}\) tensors.  In this limit, \(\boldsymbol{\epsilon}\) and \(\boldsymbol{\mu}\) become symmetric tensors, and the magneto electric coupling, \(\boldsymbol{\chi}_{\text{\tiny{E\,B}}}\) must appear with a minus sign and a transpose in \(\tilde{\boldsymbol{H}}\) versus \(\tilde{\boldsymbol{D}}\).  It is inconsistent with the Lagrangian to suppose that the static magnetoelectric coupling can have any other form.
	\par
	The constitutive relations, (\ref{tellegen-cons}), when extended to arbitrary \(\omega\), are consistent with the kinds of magnetoelectric coupling required for both moving media and Tellegen media~\cite{volume8,tellegen1948,lindell1994}.  From the point of view of the matter--field coupling terms in the Lagrangian description, there does not appear to be any contradiction in assuming that through violating both parity and time reversal symmetry, such media could be constructed in the laboratory.\\
	\par	
	This may well be true for Tellegen media, however, for real moving media there is a subtlety that means that the magnetoelectric coupling in (\ref{tellegen-cons}) does not contain all of the physics of electromagnetism interacting with a medium in motion.  As shall be shown in section~\ref{moving-medium-section}, the reservoir Lagrangian density, \(\mathscr{L}_{\text{\tiny{R}}}\) also has to be altered in this case, and it is not obvious how one would engineer a medium where the loss mechanism works in such a peculiar way.
%
%
	\subsection{Further restrictions on the constitutive relations}
	\par	
	From the above discussion, it appears that the general susceptibility tensors, (\ref{eebb-sus}--\ref{ebbe-sus}) encompass all known magnetoelectric constitutive relations, and naturally restrict the relationship between \(\boldsymbol{\chi}_{\text{\tiny{E\,B}}}\) and \(\boldsymbol{\chi}_{\text{\tiny{B\,E}}}\).  Our point of view is that this is the correct description, as the electromagnetic field may be quantised within such a formalism.  This point of view is bolstered by the fact that some of the expected restrictions from thermodynamics have also arisen along the way. 
	\par
	This formalism puts further restrictions on the susceptibilities, due to the fact that (\ref{eebb-sus}--\ref{ebbe-sus}) are related to one another.  For example \(\boldsymbol{\chi}_{\text{\tiny{E\,E}}}\) \& \(\boldsymbol{\chi}_{\text{\tiny{B\,B}}}\) together contain the same eight coupling tensors as \(\boldsymbol{\chi}_{\text{\tiny{E\,B}}}\).  Therefore it is not possible to choose the magnetoelectric coupling in a way that is independent of the value of the electric--electric and magnetic--magnetic susceptibilities.   We now proceed to work out the implications of this relationship.
	\par
	The dissipative part of the susceptibilities can be summarized as follows,
	\begin{align}
		\frac{\omega}{i\pi}\left[\boldsymbol{\chi}_{\text{\tiny{E\,E}}}-\boldsymbol{\chi}_{\text{\tiny{E\,E}}}^{\dagger}\right]&=\boldsymbol{\lambda}_{\text{\tiny{E\,E}}}\boldsymbol{\cdot}\boldsymbol{\lambda}_{\text{\tiny{E\,E}}}^{\dagger}+\boldsymbol{\lambda}_{\text{\tiny{E\,B}}}\boldsymbol{\cdot}\boldsymbol{\lambda}_{\text{\tiny{E\,B}}}^{\dagger}\nonumber\\
		\frac{\omega}{i\pi}\left[\boldsymbol{\chi}_{\text{\tiny{B\,B}}}-\boldsymbol{\chi}_{\text{\tiny{B\,B}}}^{\dagger}\right]&=\boldsymbol{\lambda}_{\text{\tiny{B\,B}}}\boldsymbol{\cdot}\boldsymbol{\lambda}_{\text{\tiny{B\,B}}}^{\dagger}+\boldsymbol{\lambda}_{\text{\tiny{B\,E}}}\boldsymbol{\cdot}\boldsymbol{\lambda}_{\text{\tiny{B\,E}}}^{\dagger}\nonumber\\
		\frac{\omega}{i\pi}\left[\boldsymbol{\chi}_{\text{\tiny{E\,B}}}-\boldsymbol{\chi}_{\text{\tiny{B\,E}}}^{\dagger}\right]&=\boldsymbol{\lambda}_{\text{\tiny{E\,E}}}\boldsymbol{\cdot}\boldsymbol{\lambda}_{\text{\tiny{B\,E}}}^{\dagger}+\boldsymbol{\lambda}_{\text{\tiny{E\,B}}}\boldsymbol{\cdot}\boldsymbol{\lambda}_{\text{\tiny{B\,B}}}^{\dagger}\nonumber\\
	\end{align}
	The problem of finding the constraints on the components of the magnetoelectric susceptibilities versus the electric--electric and magnetic--magnetic ones is now one of linear algebra.  There are four complex matrices, and we have to work out how the components resulting from multiplying them together in one way are related to multiplying them together in another way.  In appendix \ref{appendix_A} this is calculated, and the following restriction on the susceptibility tensors is obtained,
	\begin{equation}
		\left|\left[\boldsymbol{\chi}_{\text{\tiny{E\,B}}}-\boldsymbol{\chi}_{\text{\tiny{B\,E}}}^{\dagger}\right]_{ij}\right|^{2}\leq\bigg|\left[\boldsymbol{\chi}_{\text{\tiny{E\,E}}}-\boldsymbol{\chi}_{\text{\tiny{E\,E}}}^{\dagger}\right]_{ii}\bigg|\bigg|\left[\boldsymbol{\chi}_{\text{\tiny{B\,B}}}-\boldsymbol{\chi}_{\text{\tiny{B\,B}}}^{\dagger}\right]_{jj}\bigg|.\label{inequality}	
	\end{equation}
	If the Lagrangian with the coupling (\ref{all-couplings-1}--\ref{all-couplings-2}) is taken to describe the most general kind of magnetoelectric medium, then every such medium should exhibit dissipation that satisfies (\ref{inequality}) in order to be consistent with quantization.
	\par
	It is worth examining (\ref{inequality}) in a few specific cases.  In the case of chiral media, using the notation of section \ref{chiral-case}, the inequality becomes,
	\[
		\omega^{2}\Im\left[\boldsymbol{\kappa}\right]_{ij}^{2}\leq\Im\left[\boldsymbol{\chi}_{\text{\tiny{E\,E}}}\right]_{ii}\Im\left[\boldsymbol{\chi}_{\text{\tiny{B\,B}}}\right]_{jj},	
	\]
	which in the isotropic case is identical to the result of~\cite{sihvola2001} (\footnote{One should remember to translate between the difference in the forms of constitutive relations used in~\cite{sihvola2001}, and that used here, c.f.~\cite{horsley2011}}).
	\par
	For the case of magnetoelectrics in the \(\omega\to0\) limit, the inequality is neither that of~\cite{odell1963} nor~\cite{brown1968},
	\[
		\Im[\boldsymbol{\chi}_{\text{\tiny{E\,B}}}]_{ij}^{2}\leq\Im[\boldsymbol{\chi}_{\text{\tiny{E\,E}}}]_{ii}\Im[\boldsymbol{\chi}_{\text{\tiny{B\,B}}}]_{jj}
	\]
	\par
	The restriction (\ref{inequality}) affects the dissipative part of the susceptibilities at all frequencies.  Of course, this restricts the non--dissipative parts in some way as well, via the Kramers--Kronig relations.  However, for a given \emph{fixed} frequency, the Lagrangian does not appear to place restrictions on the non--dissipative parts of (\ref{eebb-sus}--\ref{ebbe-sus}).  This is contrary to the inequalities that are often used in the literature, which do not seem to fully treat dispersion or loss.
	\section{The Lagrangian for a moving medium\label{moving-medium-section}}
	\par
	It was established in section~\ref{tellegen-subsection}, that the magnetoelectric \emph{coupling} terms involving \(\boldsymbol{\alpha}_{\text{\tiny{E\,B (B\,E)}}}\) within the Lagrangian can in principle reproduce the constitutive relations for a moving medium (with the \(\boldsymbol{\beta}\) tensors equal to zero).  This section is motivated by a recent discussion regarding the existence of a frictional component to the Casimir force~\cite{leonhardt2010,pendry2010}, where it has been pointed out that if a magnetoelectric could perfectly mimic a moving medium, then the frictional force could extract work from the vacuum.  There is either no frictional force, or a magnetoelectric cannot perfectly reproduce the physics of a moving medium.  Here we argue that, from the point of view of a Lagrangian description, a magnetoelectric coupling alone is not sufficient to mimic a moving medium.  
	\par
	Suppose that in the comoving (primed) frame we have a medium that can be described via \(\boldsymbol{\epsilon}\) and \(\boldsymbol{\mu}\) alone, as was assumed in section~\ref{summary-section},
	\begin{multline}
		\mathscr{L}=\mathscr{L}_{\text{\tiny{F}}}+\mathscr{L}_{\text{\tiny{R}}}+\boldsymbol{E}^{\prime}\boldsymbol{\cdot}\int_{0}^{\infty}\boldsymbol{\alpha}_{\text{\tiny{EE}}}^{\prime}(\omega^{\prime})\boldsymbol{\cdot}\boldsymbol{X}^{\prime}_{\omega^{\prime}}d\omega^{\prime}\\
		+\boldsymbol{B}^{\prime}\boldsymbol{\cdot}\int_{0}^{\infty}\boldsymbol{\alpha}^{\prime}_{\text{\tiny{BB}}}(\omega^{\prime})\boldsymbol{\cdot}\boldsymbol{Y}^{\prime}_{\omega^{\prime}}d\omega^{\prime}\label{rest-frame-interaction}	
	\end{multline}
	Now consider the lab (unprimed) frame, where the medium is in uniform motion.  Without loss of generality, we can assume the motion is along the \(x\) axis, \(\boldsymbol{V}=V_{x}\hat{\boldsymbol{x}}\). The form of the Lagrangian density associated with the free field is unchanged in terms of the field strengths, as it is a scalar formed from \(F_{\mu\nu}F^{\mu\nu}\).  However, \(\mathscr{L}_{\text{\tiny{INT}}}\) and \(\mathscr{L}_{\text{\tiny{R}}}\) will not take the same form in terms of the oscillator amplitudes in both frames.  To find the form of the Lagrangian density in terms of lab frame quantities, we begin by transforming the field strengths,
	\begin{align}
		E_{x}^{\prime}&=E_{x}&B_{x}^{\prime}&=B_{x}\nonumber\\
		E_{y}^{\prime}&=\gamma\left(E_{y}-V_{x}B_{z}\right)&B_{y}^{\prime}&=\gamma\left(B_{y}+V_{x}E_{z}/c^{2}\right)\nonumber\\
		E_{z}^{\prime}&=\gamma\left(E_{z}+V_{x}B_{y}\right)&B_{z}^{\prime}&=\gamma\left(B_{z}-V_{x}E_{y}/c^{2}\right)\label{field-transformation}
	\end{align}
	with, \(\gamma=(1-\boldsymbol{V}^{2}/c^{2})^{-1/2}\).  Inserting these into (\ref{rest-frame-interaction}) yields an interaction Lagrangian,
	\begin{multline}
		\mathscr{L}_{\text{\tiny{INT}}}=\boldsymbol{E}\boldsymbol{\cdot}\int_{0}^{\infty}\left[\boldsymbol{\alpha}_{\text{\tiny{E\,E}}}(\omega)\boldsymbol{\cdot}\boldsymbol{X}_{\omega}+\boldsymbol{\alpha}_{\text{\tiny{E\,B}}}(\omega)\boldsymbol{\cdot}\boldsymbol{Y}_{\omega}\right]d\omega\\
		+\boldsymbol{B}\boldsymbol{\cdot}\int_{0}^{\infty}\left[\boldsymbol{\alpha}_{\text{\tiny{B\,B}}}(\omega)\boldsymbol{\cdot}\boldsymbol{Y}_{\omega}+\boldsymbol{\alpha}_{\text{\tiny{B\,E}}}(\omega)\boldsymbol{\cdot}\boldsymbol{X}_{\omega}\right]d\omega\label{lab-frame-interaction}
	\end{multline}
	with the following coupling tensors,
	\begin{align}
		\boldsymbol{\alpha}_{\text{\tiny{E\,E}}}(\omega)&=\boldsymbol{\Lambda}\boldsymbol{\cdot}\boldsymbol{\alpha}^{\prime}_{\text{\tiny{E\,E}}}(\omega)&\boldsymbol{\alpha}_{\text{\tiny{E\,B}}}(\omega)&=\gamma\boldsymbol{V}\boldsymbol{\times}\boldsymbol{\alpha}^{\prime}_{\text{\tiny{B\,B}}}(\omega)/c^{2}\nonumber\\
		\boldsymbol{\alpha}_{\text{\tiny{B\,B}}}(\omega)&=\boldsymbol{\Lambda}\boldsymbol{\cdot}\boldsymbol{\alpha}^{\prime}_{\text{\tiny{B\,B}}}(\omega)&\boldsymbol{\alpha}_{\text{\tiny{B\,E}}}(\omega)&=-\gamma\boldsymbol{V}\boldsymbol{\times}\boldsymbol{\alpha}^{\prime}_{\text{\tiny{E\,E}}}(\omega)\label{transformed-coupling-tensors}
	\end{align}
	where \(\boldsymbol{\Lambda}=\text{diag}\left(1,\gamma,\gamma\right)\).  When the medium is non--uniform in the rest frame, the coupling tensors become functions of time in the lab frame (e.g. \(\boldsymbol{\alpha}_{\text{\tiny{E\,E}}}(x^{\prime})=\boldsymbol{\alpha}_{\text{\tiny{E\,E}}}(\gamma(x-V_{x}t))\)).  Thus far the coupling tensors have been assumed to be independent of time. 
	\par
	The primes on the oscillator amplitudes and frequency have been dropped in obtaining (\ref{lab-frame-interaction}) from (\ref{rest-frame-interaction}).  This is because, as explained in the introduction, these oscillator amplitudes are an unobservable accounting device for the lost field energy.  We only have to make sure that their dynamics and coupling to the field are properly described in terms of lab frame coordinates and fields.  The amplitudes themselves cannot be observed in either frame, and so there is no useful meaning in transforming them.
	\par	
	As initially anticipated, in terms of the interaction Lagrangian, a moving medium falls into the category of a time--irreversible medium without spatial inversion symmetry (see table \ref{symmetry-table}).  In both types of inversion the velocity changes sign, which represents different medium.  Notice that the transformed interaction Lagrangian, (\ref{lab-frame-interaction}) automatically contains the Aharonov--Casher interaction~\cite{aharonov1984}, \(\boldsymbol{V}\boldsymbol{\cdot}\boldsymbol{M}\boldsymbol{\times}\boldsymbol{E}/c^{2}\), which has a subtle origin in the multipolar expansion~\cite{horsley2006}.
	\par
	Despite the familiar interaction Lagrangian, moving media are fundamentally distinct from stationary magnetoelectrics.  This is due to the behaviour of the reservoir part of the Lagrangian density.  In the rest frame this is as in (\ref{resevoir}), but with primed quantities.  However, the derivatives of the oscillator amplitudes with respect to time in the rest frame, \(\partial\boldsymbol{X}^{\prime}_{\omega^{\prime}}/\partial t^{\prime}\) \& \(\partial\boldsymbol{Y}^{\prime}_{\omega^{\prime}}/\partial t^{\prime}\), are not equal to the derivatives with respect to time in the lab frame. To describe the dynamics of the reservoir correctly in the lab frame, the time derivative must be transformed; \(\partial/\partial t^{\prime}=\gamma(\partial/\partial t + \boldsymbol{V}\boldsymbol{\cdot}\boldsymbol{\nabla})\), and the reservoir part of the Lagrangian density becomes,
	\begin{multline}
		\mathscr{L}_{\text{\tiny{R}}}=\frac{1}{2}\int_{0}^{\infty}\Bigg\{\gamma^{2}\left(\frac{\partial\boldsymbol{X}_{\omega}}{\partial t}+\left(\boldsymbol{V}\boldsymbol{\cdot}\boldsymbol{\nabla}\right)\boldsymbol{X}_{\omega}\right)^{2}\\
		+\gamma^{2}\left(\frac{\partial\boldsymbol{Y}_{\omega}}{\partial t}+\left(\boldsymbol{V}\boldsymbol{\cdot}\boldsymbol{\nabla}\right)\boldsymbol{Y}_{\omega}\right)^{2}-\omega^{2}\left({\boldsymbol{X}_{\omega}}^{2}+{\boldsymbol{Y}_{\omega}}^{2}\right)\Bigg\}d\omega\label{moving-resevoir-lab-frame}	
	\end{multline}
	where the primes are again dropped from the amplitudes and frequency, for the reason described earlier. This is not the same modification to the reservoir Lagrangian that was made in~\cite{amooshahi2009a}, although as shown below our form does produce the correct constitutive relations.
	\par
	We propose that the sum of (\ref{field_L}), (\ref{lab-frame-interaction}), \& (\ref{moving-resevoir-lab-frame}) represents the Lagrangian for the description of a medium in uniform motion, when, in the rest frame this medium can be described by the tensors \(\boldsymbol{\epsilon}\) and \(\boldsymbol{\mu}\)~\footnote{It would be interesting to consider the motion of general magnetoelectrics, for in this case there is a mixing of magnetoelectric coupling terms due to the motion.  For example, see~\cite{mackay2007}}.  An extension to non--uniform motion is possible through considering a local rest frame Lagrangian at each point in the medium, however this case is not considered here.  Notice that the key feature of (\ref{moving-resevoir-lab-frame}) is that the reservoir is fundamentally altered, \emph{even in the absence of the field}.  This is encoded within a coupling between neighbouring oscillators through the terms, \((\boldsymbol{V}\boldsymbol{\cdot}\boldsymbol{\nabla})\boldsymbol{X}_{\omega}\) \& \((\boldsymbol{V}\boldsymbol{\cdot}\boldsymbol{\nabla})\boldsymbol{Y}_{\omega}\).
	\par
	To show that \(\mathscr{L}=(\ref{field_L})+(\ref{lab-frame-interaction})+(\ref{moving-resevoir-lab-frame})\) is the correct Lagrangian, we examine the polarization and magnetization of the medium that arises from the equations of motion.  Examining (\ref{lab-frame-interaction}) shows that the polarization and magnetization are given by,
	\begin{align}
		\boldsymbol{P}&=\int_{0}^{\infty}\left[\boldsymbol{\alpha}_{\text{\tiny{E\,E}}}(\omega,\boldsymbol{x},t)\boldsymbol{\cdot}\boldsymbol{X}_{\omega}+\boldsymbol{\alpha}_{\text{\tiny{E\,B}}}(\omega,\boldsymbol{x},t)\boldsymbol{\cdot}\boldsymbol{Y}_{\omega}\right]d\omega\nonumber\\
		\boldsymbol{M}&=\int_{0}^{\infty}\left[\boldsymbol{\alpha}_{\text{\tiny{B\,B}}}(\omega,\boldsymbol{x},t)\boldsymbol{\cdot}\boldsymbol{Y}_{\omega}+\boldsymbol{\alpha}_{\text{\tiny{B\,E}}}(\omega,\boldsymbol{x},t)\boldsymbol{\cdot}\boldsymbol{X}_{\omega}\right]d\omega\label{moving-media-P-M}
	\end{align}
	To find the quantities in (\ref{moving-media-P-M}) in terms of the fields, we solve the equations of motion of the oscillators, which are now,
	\begin{align}
		\gamma^{2}\left(\frac{\partial}{\partial t}+\boldsymbol{V}\boldsymbol{\cdot}\boldsymbol{\nabla}\right)^{2}\boldsymbol{X}_{\omega}&=-\omega^{2}\boldsymbol{X}_{\omega}+\boldsymbol{\alpha}_{\text{\tiny{E\,E}}}^{T}\boldsymbol{\cdot}\boldsymbol{E}+\boldsymbol{\alpha}_{\text{\tiny{B\,E}}}^{T}\boldsymbol{\cdot}\boldsymbol{B}\nonumber\\
		\gamma^{2}\left(\frac{\partial}{\partial t}+\boldsymbol{V}\boldsymbol{\cdot}\boldsymbol{\nabla}\right)^{2}\boldsymbol{Y}_{\omega}&=-\omega^{2}\boldsymbol{Y}_{\omega}+\boldsymbol{\alpha}_{\text{\tiny{B\,B}}}^{T}\boldsymbol{\cdot}\boldsymbol{B}+\boldsymbol{\alpha}_{\text{\tiny{E\,B}}}^{T}\boldsymbol{\cdot}\boldsymbol{E}\label{moving-resevoir-dynamics}
	\end{align}
	The dynamics of (\ref{moving-resevoir-dynamics}) clearly demonstrates some kind of spatial dispersion.  Assuming a uniform medium, we write (\ref{moving-resevoir-dynamics}) in Fourier space to find \(\tilde{\boldsymbol{X}}_{\omega}\),
	\begin{multline}
		\tilde{\boldsymbol{X}}_{\omega}(\boldsymbol{k},\Omega)=\frac{\boldsymbol{\alpha}_{\text{\tiny{E\,E}}}^{T}(\omega)\boldsymbol{\cdot}\tilde{\boldsymbol{E}}(\boldsymbol{k},\Omega)+\boldsymbol{\alpha}_{\text{\tiny{B\,E}}}^{T}(\omega)\boldsymbol{\cdot}\tilde{\boldsymbol{B}}(\boldsymbol{k},\Omega)}{\left(\omega-\Omega^{\prime}-i\eta\right)\left(\omega+\Omega^{\prime}+i\eta\right)}\\
		+2\pi\gamma\left[\tilde{\boldsymbol{h}}_{\text{\tiny{X}}\omega}(\boldsymbol{k})\delta(\omega-\Omega^{\prime})+\tilde{\boldsymbol{h}}^{\star}_{\text{\tiny{X}}\omega}(-\boldsymbol{k})\delta(\omega+\Omega^{\prime})\right].\label{uniform-moving-medium}
	\end{multline}
	where \(\Omega^{\prime}=\gamma(\Omega-\boldsymbol{V}\boldsymbol{\cdot}\boldsymbol{k})\), \(\text{sgn}(\eta)=+1\) (a moving medium with loss), and the result for \(\boldsymbol{Y}_{\omega}\) is obtained from interchanging the subscripts and fields, \(E\leftrightarrow B\), \& \(X\leftrightarrow Y\).
	Note that spatial dispersion occurs in the denominator, which changes the positions of the poles.  In short: the modified dynamics of the reservoir represents the physics of the Doppler effect, which is a very special kind of spatial dispersion.  We emphasise that for this reason, this medium is not entirely equivalent to a stationary one with a magnetoelectric coupling such as that of section \ref{tellegen-subsection}.
	When (\ref{uniform-moving-medium}) is inserted in (\ref{moving-media-P-M}) along with the corresponding expression for \(\tilde{\boldsymbol{Y}_{\omega}}\), we obtain the following polarization and magnetization vectors,
	\begin{multline}
		\boldsymbol{P}=\boldsymbol{P}_{0}+\int_{\mathbb{R}^{3}}\frac{d^{3}\boldsymbol{k}}{\left(2\pi\right)^{3}}\int_{\boldsymbol{V}\boldsymbol{\cdot}\boldsymbol{k}}^{\infty}\Big\{\big[\boldsymbol{\chi}_{\text{\tiny{E\,E}}}(\omega^{\prime})\boldsymbol{\cdot}\boldsymbol{E}(\boldsymbol{k},\omega)\\
		+\boldsymbol{\chi}_{\text{\tiny{E\,B}}}(\omega^{\prime})\boldsymbol{\cdot}\boldsymbol{B}(\boldsymbol{k},\omega)\big]e^{i(\boldsymbol{k}\boldsymbol{\cdot}\boldsymbol{x}-\omega t)}+\text{{\normalsize c.c.}}\Big\}d\omega\label{moving-medium-polarization}
	\end{multline}
	and,
	\begin{multline}
		\boldsymbol{M}=\boldsymbol{M}_{0}+\int_{\mathbb{R}^{3}}\frac{d^{3}\boldsymbol{k}}{\left(2\pi\right)^{3}}\int_{\boldsymbol{V}\boldsymbol{\cdot}\boldsymbol{k}}^{\infty}\Big\{\big[\boldsymbol{\chi}_{\text{\tiny{B\,B}}}(\omega^{\prime})\boldsymbol{\cdot}\boldsymbol{B}(\boldsymbol{k},\omega)\\
		+\boldsymbol{\chi}_{\text{\tiny{B\,E}}}(\omega^{\prime})\boldsymbol{\cdot}\boldsymbol{E}(\boldsymbol{k},\omega)\big]e^{i(\boldsymbol{k}\boldsymbol{\cdot}\boldsymbol{x}-\omega t)}+\text{{\normalsize c.c.}}\Big\}d\omega\label{moving-medium-magnetization}
	\end{multline}
	where \(\omega^{\prime}=\gamma(\omega-\boldsymbol{V}\boldsymbol{\cdot}\boldsymbol{k})\), and the real and imaginary parts of the susceptibilities are related by (\ref{kramers-kronig}), with \(\omega\to\omega^{\prime}\).  The first thing to notice about (\ref{moving-medium-polarization}) \& (\ref{moving-medium-magnetization}) is that in the lab frame, some of the positive rest frame frequencies (\(\omega^{\prime}>0\)) appear as negative frequencies (\(\omega<0\)).  The response of the medium to a constant field, \(\omega^{\prime}=0\), also appears at a finite frequency in the lab, \(\omega=\boldsymbol{V}\boldsymbol{\cdot}\boldsymbol{k}\).  These peculiar features are a consequence of the Doppler effect, which has arisen from the modified reservoir dynamics encoded in (\ref{moving-resevoir-lab-frame}), and would not have occurred for any stationary magnetoelectric.  The undriven part of the polarization and magnetization also exhibits this behaviour, for example,
	\begin{multline}
		\boldsymbol{P}_{0}=\gamma\int_{\mathbb{R}^{3}}\frac{d^{3}\boldsymbol{k}}{(2\pi)^{3}}\int_{\boldsymbol{V}\boldsymbol{\cdot}\boldsymbol{k}}^{\infty}\big[\boldsymbol{\alpha}_{\text{\tiny{E\,E}}}(\omega^{\prime})\boldsymbol{\cdot}\tilde{\boldsymbol{h}}_{\text{\tiny{X}}\omega^{\prime}}(\boldsymbol{k})\\+\boldsymbol{\alpha}_{\text{\tiny{E\,B}}}(\omega^{\prime})\boldsymbol{\cdot}\tilde{\boldsymbol{h}}_{\text{\tiny{Y}}\omega^{\prime}}(\boldsymbol{k})\big]e^{i(\boldsymbol{k}\boldsymbol{\cdot}\boldsymbol{x}-\omega t)}d\omega
		+\text{\normalsize c.c.}
	\end{multline}
	The susceptibilities are given in terms of the coupling tensors as in section \ref{tellegen-subsection}.  Inserting (\ref{transformed-coupling-tensors}) into (\ref{eebb-sus}--\ref{ebbe-sus}), and identifying the rest frame susceptibilities according to (\ref{coupling-susceptibility}), we find the following transformation formulae,
	\begin{align*}
		\boldsymbol{\chi}_{\text{\tiny{E\,E}}}&=\boldsymbol{\Lambda}\boldsymbol{\cdot}\boldsymbol{\chi}_{\text{\tiny{E\,E}}}^{\prime}\boldsymbol{\cdot}\boldsymbol{\Lambda}-\frac{\gamma^{2}}{c^{2}}\frac{\boldsymbol{V}}{c}\boldsymbol{\times}\boldsymbol{\chi}_{\text{\tiny{B\,B}}}^{\prime}\boldsymbol{\times}\frac{\boldsymbol{V}}{c}\\
		\boldsymbol{\chi}_{\text{\tiny{E\,B}}}&=\gamma\left[\boldsymbol{\Lambda}\boldsymbol{\cdot}\boldsymbol{\chi}_{\text{\tiny{E\,E}}}^{\prime}\boldsymbol{\times}\boldsymbol{V}+\frac{1}{c^{2}}\boldsymbol{V}\boldsymbol{\times}\boldsymbol{\chi}_{\text{\tiny{B\,B}}}^{\prime}\boldsymbol{\cdot}\boldsymbol{\Lambda}\right]\\
		\boldsymbol{\chi}_{\text{\tiny{B\,B}}}&=\boldsymbol{\Lambda}\boldsymbol{\cdot}\boldsymbol{\chi}^{\prime}_{\text{\tiny{B\,B}}}\boldsymbol{\cdot}\boldsymbol{\Lambda}-\gamma^{2}\boldsymbol{V}\times\boldsymbol{\chi}_{\text{\tiny{E\,E}}}^{\prime}\boldsymbol{\times}\boldsymbol{V}\\
		\boldsymbol{\chi}_{\text{\tiny{B\,E}}}&=-\gamma\left[\boldsymbol{V}\boldsymbol{\times}\boldsymbol{\chi}_{\text{\tiny{E\,E}}}^{\prime}\boldsymbol{\cdot}\boldsymbol{\Lambda}+\frac{1}{c^{2}}\boldsymbol{\Lambda}\boldsymbol{\cdot}\boldsymbol{\chi}_{\text{\tiny{B\,B}}}^{\prime}\boldsymbol{\times}\boldsymbol{V}\right]
	\end{align*}
	which reduce to the well known first order in \(\boldsymbol{V}/c\) results when the medium is isotropic (\(\boldsymbol{\chi}_{\text{\tiny{E\,E}}}^{\prime}=\boldsymbol{\mathbb{1}}_{3}\chi_{\text{\tiny{E\,E}}}^{\prime}\) \& \(\boldsymbol{\chi}_{\text{\tiny{B\,B}}}^{\prime}=\boldsymbol{\mathbb{1}}_{3}\chi_{\text{\tiny{B\,B}}}^{\prime}\))~\cite{volume8},
	\begin{align}
		\boldsymbol{D}(\boldsymbol{k},\omega)&=\epsilon^{\prime}(\omega^{\prime})\boldsymbol{E}(\boldsymbol{k},\omega)+({n^{\prime}}^{2}(\omega^{\prime})-1)\frac{\boldsymbol{V}\boldsymbol{\times}\boldsymbol{H}(\boldsymbol{k},\omega)}{c^{2}}\nonumber\\
		\boldsymbol{B}(\boldsymbol{k},\omega)&=\mu^{\prime}(\omega^{\prime})\boldsymbol{H}(\boldsymbol{k},\omega)-({n^{\prime}}^{2}(\omega^{\prime})-1)\frac{\boldsymbol{V}\boldsymbol{\times}\boldsymbol{E}(\boldsymbol{k},\omega)}{c^{2}}
	\end{align}
	where, \(\epsilon^{\prime}(\omega^{\prime})=\epsilon_{0}+\chi_{\text{\tiny{E\,E}}}^{\prime}(\omega^{\prime})\), \({\mu^{\prime}}^{-1}(\omega^{\prime})=\mu_{0}^{-1}-\chi_{\text{\tiny{B\,B}}}^{\prime}(\omega^{\prime})\), and \({n^{\prime}}^{2}(\omega^{\prime})/c^{2}=\epsilon^{\prime}(\omega^{\prime}){\mu^{\prime}}(\omega^{\prime})\).  Therefore, the Lagrangian density, \(\mathscr{L}=(\ref{field_L})+(\ref{lab-frame-interaction})+(\ref{moving-resevoir-lab-frame})\) reproduces the correct macroscopic Maxwell equations for a medium in motion with any \(\boldsymbol{\epsilon}\) and \(\boldsymbol{\mu}\) that satisfy the Kramers--Kronig relations.
	\par
	From the point of view of the quantization of macroscopic QED in terms of a ficticous bath of oscillators, it may seem obvious that a moving medium should not be equivalent to a stationary magnetoelectric: a harmonic oscillator is not a relativistically invariant system.  However, if we do not utilise the bath of oscillators in the quantization, then it is not obvious that the two are inequivalent.  In regimes where dispersion is negligible, the constitutive equations take the same form, and this apparent equivalence has led to a paradox in the theory of the Casimir effect~\cite{leonhardt2010}, which we have hopefully shed some light on.
%
%
	\section{Conclusions}
	\par
	We have shown that a natural generalization of the Lagrangian in~\cite{philbin2010} provides a general description of magnetoelectrics, and media exhibiting Hermitian \(\boldsymbol{\epsilon}\) \& \(\boldsymbol{\mu}\) tensors.  It has also been established that this generalization is only consistent with quantization (i.e. the existence of a Hamiltonian) with materials where the magnetoelectric coupling satisfies (\ref{inequality}), which reproduces some inequalities previously derived from different physical arguments.  We propose (\ref{inequality}) to be an accurate restriction on the magnitude of the magnetoelectric coupling.
	\par
	It has also been shown that as far as the Lagrangian description is concerned, moving media are not equivalent to stationary magnetoelectrics.  The coupling between the field and the reservoir can indeed be the same in the two cases, however, the reservoir must be represented by (\ref{moving-resevoir-lab-frame}) rather than (\ref{resevoir}) to fully account for the physics of the Doppler effect.
	\acknowledgements
        This research is supported by an EPSRC postdoctoral fellowship award.  I thank Tom Philbin for introducing me to the theory of macroscopic quantum electromagnetism, taking the time to point out and explain the problem of negative frequencies in `quantum friction', and for the many helpful discussions and suggestions that led to this work.  Thanks go to the referee, who's comments greatly improved the manuscript.
	\bibliography{refs}
	\appendix
	%
%
%
%
	\section{A derivation of the magnetoelectric inequality\label{appendix_A}}
	\par
	Suppose we have four \(3\times3\) matrices, \(\boldsymbol{\alpha}\), \(\boldsymbol{\beta}\), \(\boldsymbol{\gamma}\), \& \(\boldsymbol{\delta}\), all containing complex entries.  We write these matrices in a compressed notation as follows,
	\[
		\boldsymbol{\alpha}=\left(\begin{matrix}
			\alpha_{11}&\alpha_{12}&\alpha_{13}\\
			\alpha_{21}&\alpha_{22}&\alpha_{23}\\
			\alpha_{31}&\alpha_{32}&\alpha_{33}		
		\end{matrix}\right)=\left(\begin{matrix}
			\boldsymbol{\alpha}_{1}\\
			\boldsymbol{\alpha}_{2}\\
			\boldsymbol{\alpha}_{3}		
		\end{matrix}\right),
	\]
	where, \(\boldsymbol{\alpha}_{i}=\left(\alpha_{i1},\alpha_{i2},\alpha_{i3}\right)\).  We now seek to find the relationship between the following matrix products,
	\begin{align*}
		\boldsymbol{\lambda}&=\boldsymbol{\alpha}\boldsymbol{\cdot}\boldsymbol{\alpha}^{\dagger}+\boldsymbol{\gamma}\boldsymbol{\cdot}\boldsymbol{\gamma}^{\dagger}\\
		\boldsymbol{\nu}&=\boldsymbol{\beta}\boldsymbol{\cdot}\boldsymbol{\beta}^{\dagger}+\boldsymbol{\delta}\boldsymbol{\cdot}\boldsymbol{\delta}^{\dagger}\\
		\boldsymbol{\sigma}&=\boldsymbol{\alpha}\boldsymbol{\cdot}\boldsymbol{\delta}^{\dagger}+\boldsymbol{\gamma}\boldsymbol{\cdot}\boldsymbol{\beta}^{\dagger}
	\end{align*}
	Expanding out these matrix products, we find that their elements can be written as follows,
	\begin{align*}
		\lambda_{ij}&=\boldsymbol{\alpha}_{i}\boldsymbol{\cdot}\boldsymbol{\alpha}_{j}^{\star}+\boldsymbol{\gamma}_{i}\boldsymbol{\cdot}\boldsymbol{\gamma}_{j}^{\star}\\
		\nu_{ij}&=\boldsymbol{\beta}_{i}\boldsymbol{\cdot}\boldsymbol{\beta}_{j}^{\star}+\boldsymbol{\delta}_{i}\boldsymbol{\cdot}\boldsymbol{\delta}_{j}^{\star}\\
		\sigma_{ij}&=\boldsymbol{\alpha}_{i}\boldsymbol{\cdot}\boldsymbol{\delta}_{j}^{\star}+\boldsymbol{\gamma}_{i}\boldsymbol{\cdot}\boldsymbol{\beta}_{j}^{\star}
	\end{align*}
	We now examine the matrix element, \(\sigma_{ij}\) and look to write it in terms of the diagonal terms, \(\lambda_{ii}\) \& \(\nu_{jj}\).  Multiplying together both these diagonal terms, we obtain,
	\begin{equation}
		\lambda_{ii}\nu_{jj}=|\boldsymbol{\alpha}_{i}|^{2}|\boldsymbol{\beta}_{j}|^{2}+|\boldsymbol{\alpha}_{i}|^{2}|\boldsymbol{\delta}_{j}|^{2}+|\boldsymbol{\gamma}_{i}|^{2}|\boldsymbol{\beta}_{j}|^{2}+|\boldsymbol{\gamma}_{i}|^{2}|\boldsymbol{\delta}_{j}|^{2}\label{matrix-product}
	\end{equation}
	We compare this with the absolute square of the element, \(\sigma_{ij}\),
	\begin{equation}
		\left|\sigma_{ij}\right|^{2}=\left|\boldsymbol{\alpha}_{i}\boldsymbol{\cdot}\boldsymbol{\delta}_{j}^{\star}\right|^{2}+|\boldsymbol{\gamma}_{i}\boldsymbol{\cdot}\boldsymbol{\beta}_{j}^{\star}|^{2}+2\Re\left[(\boldsymbol{\alpha}_{i}\boldsymbol{\cdot}\boldsymbol{\delta}_{j}^{\star})(\boldsymbol{\gamma}_{i}\boldsymbol{\cdot}\boldsymbol{\beta}_{j}^{\star})\right]\label{squared-element}
	\end{equation}
	Applying the Cauchy--Schwarz inequality~\cite{byron1992}, \(|\boldsymbol{x}|^{2}|\boldsymbol{y}|^{2}\geq|\boldsymbol{x}\boldsymbol{\cdot}\boldsymbol{y}|^{2}\), to (\ref{squared-element}) we thus obtain,
	\[
		\left|\sigma_{ij}\right|^{2}\leq\left|\boldsymbol{\alpha}_{i}|^{2}|\boldsymbol{\delta}_{j}\right|^{2}+|\boldsymbol{\gamma}_{i}|^{2}|\boldsymbol{\beta}_{j}|^{2}+2|\boldsymbol{\alpha}_{i}||\boldsymbol{\delta}_{j}||\boldsymbol{\gamma}_{i}||\boldsymbol{\beta}_{j}|
	\]
	or, from (\ref{matrix-product}),
	\[
		\left|\sigma_{ij}\right|^{2}\leq\lambda_{ii}\nu_{jj}-(|\boldsymbol{\alpha}_{i}||\boldsymbol{\beta}_{j}|-|\boldsymbol{\gamma}_{i}||\boldsymbol{\delta}_{j}|)^{2}
	\]
	So finally we find the following inequality must be satisfied by the matrix elements,
	\[
			\left|\sigma_{ij}\right|^{2}\leq\lambda_{ii}\nu_{jj}.
	\]
	
\end{document}